\documentclass[pdflatex,sn-mathphys-num]{sn-jnl}

\usepackage{graphicx}%
\usepackage{multirow}%
\usepackage{amsmath,amssymb,amsfonts}%
\usepackage{amsthm}%
\usepackage{mathrsfs}%
\usepackage[title]{appendix}%
\usepackage{xcolor}%
\usepackage{textcomp}%
\usepackage{manyfoot}%
\usepackage{booktabs}%
\usepackage{algorithm}%
\usepackage{algorithmicx}%
\usepackage{algpseudocode}%
\usepackage{listings}%
\usepackage{tabularx} 
\usepackage{multirow} 
\usepackage{array} 
\usepackage{float}
\usepackage{caption} 
\usepackage{pdflscape} 
\usepackage{placeins} 
\usepackage{graphicx}
\usepackage{threeparttable}
\usepackage{lineno}

\usepackage{color}
\usepackage{xcolor}
\definecolor{gold}{RGB}{217, 165, 33} 
\definecolor{silver}{RGB}{153, 153, 153} 
\definecolor{bronze}{RGB}{184, 110, 31} 
\definecolor{green}{RGB}{0, 164, 74} 
\definecolor{red}{RGB}{159, 45, 5} 

\theoremstyle{thmstyleone}%
\theoremstyle{thmstyletwo}%

\theoremstyle{thmstylethree}%

\begin{document}

\title[Article Title]{Genotype-to-Phenotype Prediction in Rice with High-Dimensional Nonlinear Features}


\author[1]{\fnm{Zeyuan} \sur{Zhou}}\email{2413393093@st.gxu.edu.cn}\equalcont{These authors contributed equally to this work.}
\author[1]{\fnm{Siyuan} \sur{Chen}}\email{seaward901@gmail.com}\equalcont{These authors contributed equally to this work.}
\author[1]{\fnm{Xinzhang} \sur{Wu}}\email{xwu@gxu.edu.cn}
\author*[1]{\fnm{Jisen} \sur{Zhang}}\email{zjisen@126.com}
\author*[1]{\fnm{Yunxuan} \sur{Dong}}\email{dongyx@gxu.edu.cn}

\affil[1]{\orgname{Guangxi University}}


\abstract{Genotype-to-Phenotype prediction can promote advances in modern genomic research and crop improvement, guiding precision breeding and genomic selection. However, high-dimensional nonlinear features often hinder the accuracy of genotype-to-phenotype prediction by increasing computational complexity. The challenge also limits the predictive accuracy of traditional approaches. Therefore, effective solutions are needed to improve the accuracy of genotype-to-phenotype prediction. In our paper, we propose MLFformer. MLFformer is a Transformer-based architecture that incorporates the Fast Attention mechanism and a multilayer perceptron module to handle high-dimensional nonlinear features. In MLFformer, the Fast Attention mechanism is utilized to handle computational complexity and enhance processing efficiency. In addition, the MLP structure further captures high-dimensional nonlinear features. Through experiments, the results show that MLFformer reduces the average MAPE by 7.73\% compared to the vanilla Transformer. In univariate and multivariate prediction scenarios, MLFformer achieves the best predictive performance among all compared models.
}




\maketitle

\section{Introduction}\label{sec1}
    The relationships among Genotype-to-Phenotype prediction (G2P) play a central role in advancing agriculture, particularly in crop breeding and trait improvement \citep{fernandez2015sol}. By enabling accurate phenotype prediction, genotypic data allowed researchers to identify genetic variations associated with critical traits such as yield, disease resistance, and stress tolerance \citep{crossa2024machine}\citep{borrill2019applying}. The application of genotypic data in phenotype prediction facilitated the optimization of breeding strategies and addressed the growing need for sustainable agricultural practices \citep{tuggle2024current}.

    Recent advancements in computer science, particularly in deep learning, demonstrated the ability to process high-dimensional nonlinear features across various domains\citep{li2020deep}. The integration of deep learning methods into agriculture enhanced G2P prediction by uncovering intricate patterns within genomic data \citep{yan2023machine}. The developments improved the precision of trait prediction and supported advancements in crop improvement and sustainable agricultural practices.

    Conventional G2P prediction methods, including statistical models such as GWAS, BLUP, and linear regression, faced challenges when analyzing high-dimensional nonlinear features \citep{schulthess2018advantages}. Moreover, the approaches struggled to model high-dimensional nonlinear features and often lacked scalability for datasets with multiple phenotypes \citep{li2016two}. Furthermore, traditional methods exhibited sensitivity to missing data, which limited their applicability in practical agricultural applications \citep{lyngdoh2022prediction}.

    Then, deep learning-based approaches demonstrate potential in G2P prediction \citep{zou2019primer}. However, challenges arose when these methods were applied to genomic datasets. High-dimensional nonlinear features often resulted in overfitting, creating difficulties in capturing meaningful patterns \citep{yin2020kaml}. Although some progress is made, a systematic exploration of the issue remains insufficient, emphasizing the need for robust and scalable solutions.

    Therefore, the key question is how to design a robust and scalable deep-learning architecture that captures high-dimensional nonlinear patterns and provides accurate G2P predictions. 

    To begin with, early applications of deep-learning architectures for G2P prediction introduced methods such as convolutional neural networks and recurrent neural networks \citep{liu2020application}. The models offered significant advantages in extracting spatial and sequential features from genomic data, enhancing predictive performance \citep{dias2019artificial}. However, the methods faced limitations in processing high-dimensional datasets, often struggling with computational inefficiency and the inability to fully capture high-dimensional nonlinear features \citep{dalla2024nucleotide}. Therefore, a model capable of handling high-dimensional nonlinear features while reducing computational complexity is required.

    As research progressed, subsequent advancements addressed some of the issues by adopting Transformer-based architectures, which utilized self-attention mechanisms to model long-range dependencies within genomic data \citep{lam2024large}. The approaches improved scalability and efficiency compared to earlier deep-learning methods \citep{liu2024pdllms}. However, high-dimensional nonlinear features posed new challenges \citep{ayesha2020overview}. The computational complexity and multicollinearity in high-dimensional nonlinear features made it difficult to model intricate patterns \citep{liu2022surrogate}. The challenges highlighted the need for further innovation.

    To address the challenge of low G2P prediction accuracy caused by the difficulty in handling high-dimensional nonlinear features, our study proposes MLFformer. The Transformer framework captures dependencies within high-dimensional nonlinear features, which  can promote advances in accurate predictions. Furthermore, the Fast Attention mechanism reduces computational complexity and enhances processing, making it suitable for large datasets \citep{choromanski2020rethinking}. In addition, the multilayer perceptron (MLP) structure captures high-dimensional nonlinear features. The key structures of MLFformer demonstrate its potential in handling high-dimensional nonlinear features and achieving accurate G2P predictions.

    Given the challenge, our research offers several contributions to the G2P prediction field. First, the MLFformer architecture introduces algorithmic innovations, such as the integration of Fast Attention mechanism and MLP structures \citep{choromanski2020rethinking}\citep{pinkus1999approximation}. Second, our study evaluates the model's predictive accuracy across univariate and multivariate prediction scenarios using multiple metrics. The evaluation validates the model's generalization ability and accuracy in G2P prediction. The model demonstrates advantages in G2P prediction through systematic experiments comparing it with multiple compared models. Overall, this study provides a distinct contribution to the intersection of agriculture and computer science. Our research advances G2P prediction methods and contributes to addressing key challenges in the field.

    The subsequent sections are organized as follows: The Literature Review section systematically reviews prior research on G2P prediction and the application of deep learning in agriculture. Then, the Methods section describes the MLFformer model and experimental procedures in detail. The Experimental Results and Discussion section presents the core experimental results and comparative analysis. Finally, the Conclusion section summarizes the findings and provides perspectives for future research directions.

    \begin{table}[h]
    \centering
    \caption{Nomenclature}
    \begin{tabularx}{\textwidth}{|lX lX|}
    \hline

        \( Q \)         & Query matrix in self-attention &
        \( K \)         & Key matrix in self-attention \\
        \( V \)         & Value matrix in self-attention &
        \( D \)         & Dimensionality of the keys \\
        \( L \)         & Sequence length &
        \( Q' \)        & Random feature map projection of \( Q \) \\
        \( K' \)        & Random feature map projection of \( K \) &
        \( {1}_L \)     & A vector of ones with size \( L \), used for normalization in the attention calculation \\
        \( x \)         & Input feature vector &
        \( z \)         & Embedded features \\
        \( W_e \)       & Weight matrix of the embedding layer &
        \( b_e \)       & Bias vector of the embedding layer \\
        \( y \)         & Output of the layer &
        \( v \)         & Input of the layer \\
        \(\Theta\)      & Trainable parameters &
        \( \mu \)       & Mean of the features \\
        \( \sigma^2 \)  & Variance of the features &
        \( \epsilon \)  & Small constant added for numerical stability \\
        \( \gamma \)    & Learnable scaling parameter &
        \( \beta \)     & Learnable shifting parameter \\
        \( W_1 \)       & Weight matrix of the first linear layer in MLP &
        \( W_2 \)       & Weight matrix of the second linear layer in MLP \\
        \( b_1 \)       & Bias vector of the first linear layer in MLP &
        \( b_2 \)       & Bias vector of the second linear layer in MLP \\
        \( \mathbf{y}_{mw} \) & Observed phenotype values for sample \( m \) and phenotype \( w \) &
        \( \hat{\mathbf{y}}_{mw} \) & Predicted phenotype values for sample \( m \) and phenotype \( w \) \\
        \( \overline{\mathbf{y}}_w \) & Mean of observed phenotype values for phenotype \( w \) &
        \( d \)         & Dimensionality of the model's hidden states \\
        \( d_l \)       & Dimensionality of the input feature vector &
        \( d_e \)       & Dimensionality of the target embedding vector \\
        \( h \)         & Number of attention heads in the multi-head attention mechanism &
        \( l \)         & Depth of the model, influencing its capacity \\
        \( p \)         & Dropout rate, preventing overfitting during training &
        \( \eta \)      & Learning rate, controlling the step size during optimization \\
        \( b \)         & Batch size of the input tensor &
        \( c \)         & Feature dimension of the input tensor \\
    \hline
    \end{tabularx}
    \end{table}

\section{Literature Review}\label{sec2}

Over the past decades, G2P research in agriculture has undergone significant advancements, moving from traditional genetic approaches to the use of high-throughput sequencing technologies \citep{tuggle2022agricultural}. Traditional methods, such as Mendelian analysis, focused on identifying inheritance patterns of specific traits. With the advent of next-generation sequencing, the volume of available genotype data grew rapidly, leading to an increasing interest in developing computational models that connect genotypic data with phenotypic predictions \cite{tuggle2022agricultural}. This shift reflected the growing complexity of phenotypic traits and the need to extract meaningful insights from large-scale genetic datasets \cite{lopez2023leveraging}.
    
    In the early stages, statistical approaches, such as Best Linear Unbiased Prediction and genome-wide association studies, and machine learning methods, such as regression and random forests, were widely used in G2P prediction \citep{wu2016joint}. These methods provide tools to process genotype data and make predictions about phenotypes. However, challenges arose when dealing with high-dimensional nonlinear features. For example, the methods were often limited in their ability to handle missing values or fully utilize the richness of modern genetic datasets \citep{liu2018wind}. As a result, new approaches suitable for handling high-dimensional nonlinear features became necessary.
    
    Over time, deep learning methods, such as convolutional neural networks, recurrent neural networks, and multilayer perceptrons, were applied to G2P prediction tasks \citep{li2024trg2p}. These models are capable of learning features directly from data, without the need for manual feature engineering. In applications such as crop genomic prediction and genomic selection, these methods were used to capture high-dimensional nonlinear features. However, traditional deep learning methods struggled to capture long-range dependencies in high-dimensional nonlinear features \citep{chen2023long}. The limitation reduces their ability to model nonlinear interactions, which are critical for G2P prediction.
    
    Subsequently, the Transformer framework began to be applied to handle high-dimensional nonlinear features \citep{hu2023high}. Transformer architectures, originally develop for natural language processing, gain attention in bioinformatics due to their ability to model long-range dependencies in sequence data. Applications such as RNA secondary structure prediction and genome sequence analysis demonstrated the use of attention mechanisms for processing sequential data \citep{dalla2024nucleotide}. While Transformers show effectiveness in these areas, their application in G2P prediction, particularly for high-dimensional nonlinear features, remained relatively unexplored. Existing research provided limit evidence on how well Transformers performed in modeling relationships in high-dimensional nonlinear features, leaving an open area for further investigation.
    
    To further improve G2P prediction accuracy, our study addresses the challenges of handling high-dimensional nonlinear features in G2P prediction. Traditional methods struggle to model these features, which reduces predictive accuracy. The study introduces the Transformer architecture, Fast Attention mechanism \citep{choromanski2020rethinking}, and multilayer perceptron structure to improve prediction performance. The Transformer extracts dependencies in genomic data, Fast Attention reduces computational complexity, and the multilayer perceptron models nonlinear relationships. The innovations enhance the applicability of Transformer-based models in G2P prediction. The innovations in our study provide a new and reliable solution for handling high-dimensional nonlinear features and G2P prediction.

\section{Methods}\label{sec3}
    \subsection{Architecture of MLFformer}
\begin{figure}[h]\centering
	\includegraphics[width=\linewidth]{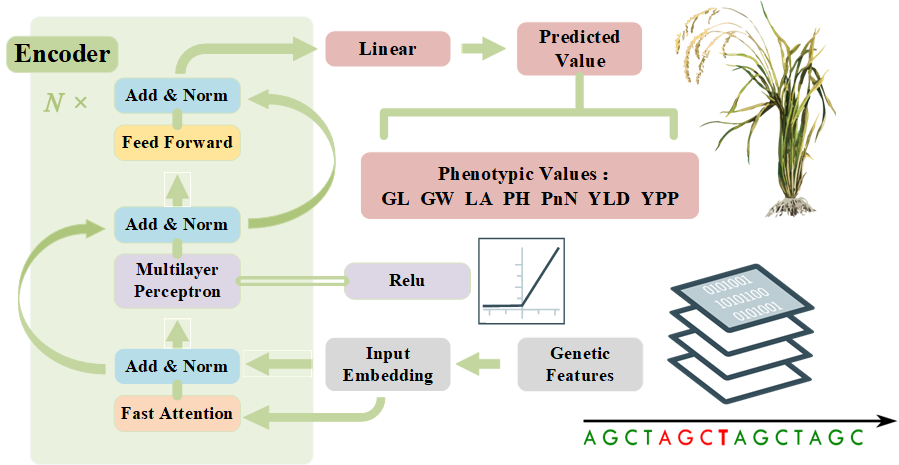}
	\caption{The MLFformer model architecture.} 
    \label{MLFformer}
    \end{figure}

    Our MLFformer model is designed to address high-dimensional nonlinear features and improve G2P prediction. The Fast Attention mechanism enhances computational efficiency when processing genomic data. Multi-Head Attention captures dependencies between high-dimensional nonlinear features. Layer Normalization stabilizes the training process by normalizing inputs across layers. The Multilayer Perceptron models complex relationships within high-dimensional nonlinear features. The Embedding Layer structures the input data for efficient processing. Residual Connections preserve critical information and prevent gradient vanishing. Each component contributes to accurate predictions in genomic data characterized by high-dimensional nonlinear features. The complete architecture of the MLFformer model is illustrated in Fig.~\ref{MLFformer}.

    \subsubsection{Fast Attention Mechanism for High-Dimensional Nonlinear Features Processing}
            In traditional Transformers, self-attention mechanisms are used to model dependencies across sequence elements by calculating a weighted representation of the input sequence. The mechanism operates based on the formula:
            \begin{align}
                \text{Attention}(Q, K, V) = \text{softmax}\left(\frac{QK^\top}{\sqrt{D}}\right)V
            \end{align}

            where $Q$, $K$, and $V$ represent the query, key, and value matrices, and $D$ is the dimensionality of the keys. This mechanism requires the computation of a complete attention matrix, which scales quadratically with the sequence length $L$. While effective, the quadratic complexity leads to high computational and memory requirements, making the approach impractical for long sequences or high-dimensional genomic datasets.
    
            Traditional self-attention mechanisms face challenges when applied to G2P prediction tasks. The high-dimensional nonlinear features pose scalability issues for quadratic complexity, making it difficult to process large-scale data. 
    
            The Fast Attention mechanism implemented in our study addresses the limitations of traditional attention mechanisms by utilizing principles from the FAVOR+ framework \citep{choromanski2020rethinking}. This method approximates the softmax kernel using random feature maps, enabling linear complexity for both computation and memory. The attention computation is reformulated as:
            \begin{align}
                \text{Attention}(Q, K, V) \approx \text{diag}(Q' (K'^\top {1}_L))^{-1} Q' (K'^\top V)
            \end{align}
            
            where $Q'$ and $K'$ are the random feature map projections of $Q$ and $K$. \({1}_L\) represents a vector of ones with a length of \(L\), where \(L\) denotes the sequence length. This formulation avoids the explicit computation of the full attention matrix, significantly improving scalability. Additionally, the flexibility to adopt various kernel functions enhances the adaptability of the mechanism to different data types.
    
            The Fast Attention mechanism offers several advantages for G2P prediction tasks. The linear complexity enables the processing of high-dimensional genomic data and long sequences, which are common in G2P studies. The use of random feature projections facilitates the modeling of complex nonlinear relationships inherent in G2P mapping. 
    
            The integration of Fast Attention mechanism within the MLFformer architecture significantly enhances its effectiveness in G2P prediction. The mechanism captures intricate genotype-to-phenotype relationships by handling long-range dependencies and nonlinear interactions. The reduction in computational costs ensures that the model remains practical for large-scale genomic datasets. Additionally, the robustness provided by the random feature approximation minimizes sensitivity to data sparsity or noise, which are common in genomic datasets. This innovative approach bridges the gap between computational efficiency and biological relevance, making the model well-suited for agricultural genomic applications.
        \subsubsection{Multi-Head Attention Mechanism for Capturing Dependencies}
            The Attention module is designed to capture dependencies among features in input data, using a multi-head attention mechanism as its core. The input tensor, with dimensions $(b, L, c)$, represents batch size $b$, sequence length $L$, and feature dimension $c$. Linear transformations are applied to project the input tensor into $Q$, $K$, and $V$, each partitioned into $h \times \frac{d}{h}$ subspaces. The Fast Attention mechanism is used to calculate the dependencies among features, and the resulting representations are processed through a linear output layer. The design efficiently extracts long-range dependencies from high-dimensional input data, enabling effective representation of input features for subsequent computations.
    
            The multi-head attention mechanism enables the Attention module to capture feature dependencies from multiple perspectives. By computing independent attention heads in parallel, the module examines different patterns in the relationships among high-dimensional nonlinear features. Each attention head works in its designated subspace, allowing for the detection of complex interactions within high-dimensional nonlinear features. In G2P prediction tasks, where high-dimensional nonlinear features often exhibit intricate relationships with phenotypes, the multi-head attention mechanism enhances the module's ability to extract meaningful patterns. This ability to simultaneously capture local and global dependencies makes the Attention module suitable for high-dimensional datasets.
    
            The Attention module incorporates the Fast Attention mechanism to optimize computational efficiency. This mechanism replaces traditional attention calculations with kernel-based random feature mapping, reducing computational complexity from quadratic to linear. Query and Key representations are projected into a lower-dimensional space using kernel functions, followed by normalization to stabilize computations. The attention weights are then computed efficiently through matrix multiplication, enabling the module to handle large-scale data without excessive resource consumption. This integration ensures that the module remains computationally feasible for applications involving high-dimensional genomic data.
    
            The Attention module is well-suited for G2P prediction due to its ability to address the nonlinear relationships and long-range dependencies inherent in genomic data. The multi-head attention mechanism captures high-dimensional nonlinear features, while the Fast Attention mechanism ensures scalability and efficiency. The attributes allow the Attention module to process complex data. The adaptability supports the module's consistent performance in capturing the high-dimensional nonlinear features, making it a valuable tool for genomic prediction tasks.
    
        \subsubsection{Layer Normalization for Genotype-to-Phenotype Model Training}
            To normalize the outputs of neurons in each layer and improve the stability and efficiency of training in G2P prediction, we employ Layer Normalization. The formula is shown below:
            \begin{align}
                \text{LN}(x) = \frac{x - \mu}{\sqrt{\sigma^2 + \epsilon}} \cdot \gamma + \beta
            \end{align}
            
            where \(x\) represents the input feature vector, \(\mu\) and \(\sigma^2\) denote the mean and variance of the features, respectively. \(\epsilon\) is a small constant added for numerical stability, and \(\gamma\) and \(\beta\) are learnable parameters used to scale and shift the normalized output.

            Layer Normalization normalizes the input feature vector across its feature dimensions by computing the mean \(\mu\) and variance \(\sigma^2\) for each feature independently. Unlike Batch Normalization, it operates at the feature level rather than across a batch, making it independent of batch size. By centering and scaling the input features within each layer, Layer Normalization reduces internal covariate shift, stabilizing and accelerating the convergence of gradient-based learning. This method is particularly effective in Transformer-based architectures due to its ability to handle sequences and high-dimensional data.

            In the context of genotype-to-phenotype (G2P) prediction, Layer Normalization is crucial for processing high-dimensional nonlinear features. By normalizing input features at each layer, it reduces variations among high-dimensional nonlinear features and ensures stable gradient propagation during training. This enhances the model's ability to learn complex nonlinear relationships between genotypes and phenotypes. Additionally, it improves robustness in handling noisy or imbalanced genomic datasets, contributing to more accurate phenotype predictions.

        \subsubsection{Multilayer Perceptron for Modeling High-Dimensional Nonlinear Features}
            To capture the complex nonlinear interactions in genomic data, the multilayer perceptron structure is applied after the attention module. The mathematical operation of the MLP is represented as:

            \begin{align}
                \text{MLP}(x) = \text{Linear}_2(\text{Dropout}(\text{ReLU}(\text{Linear}_1(x))))
            \end{align}
            
            where \(\text{Linear}_1(x) = W_1 \cdot x + b_1\) and \(\text{Linear}_2(x) = W_2 \cdot x + b_2\). Here, \(W_1\) and \(W_2\) are the weight matrices of the two linear layers, and \(b_1\) and \(b_2\) are the corresponding bias terms. The ReLU activation function is defined as:
            
            \begin{align}
                 \text{ReLU}(x) = \max(0, x)
            \end{align}
            
            where \(\max(0, x)\) ensures that negative input values are replaced with zero while retaining positive values. This activation function introduces nonlinearity, enabling the model to learn complex patterns in the data.

            The MLP structure consists of two linear transformations, a ReLU activation function, and a dropout layer. The first linear transformation reduces the dimensionality of the input, extracting critical features from high-dimensional genomic data. The ReLU activation applies a nonlinear transformation, allowing the model to capture intricate relationships between genomic markers and phenotypes. The dropout layer enhances generalization by randomly deactivating a subset of neurons during training, mitigating overfitting. The second linear transformation restores the dimensionality to its original size, ensuring compatibility with subsequent model layers.

            In the context of G2P prediction, the MLP structure is designed to model the high-dimensional nonlinear features. The combination of dimensionality reduction, nonlinear activation, and dropout improves the model's ability to generalize across diverse phenotypes and datasets. The structure balances computational with representational capacity, making it suitable for high-dimensional nonlinear features.

        \subsubsection{Embedding Layer for Feature Representation}
            To map the high-dimensional nonlinear features into a fixed-dimensional space that can be processed by the model, an embedding layer is introduced. The embedding layer projects the input features through a linear transformation into the target dimension, providing a unified input representation for subsequent Transformer modules. The formula is as follows:
            \begin{align}
                \mathbf{z} = \mathbf{W}_e \cdot \mathbf{x} + \mathbf{b}_e
            \end{align}
            where $\mathbf{z} \in \mathbb{R}^{d_e}$ represents the embedded features, $\mathbf{x} \in \mathbb{R}^{d_l}$ is the input feature vector, and $\mathbf{W}_e \in \mathbb{R}^{d_e \times d_l}$ and $\mathbf{b}_e \in \mathbb{R}^{d_e}$ are the weight matrix and bias vector of the embedding layer, respectively. $d_l$ denotes the input feature dimension, and $d_e$ denotes the target embedding dimension.

            The embedding layer is a linear mapping that learns the weight matrix $\mathbf{W}_e$ and bias vector $\mathbf{b}_e$ to project and compress the input data into a dimension suitable for the subsequent layers. In G2P prediction tasks, where the input data is typically high-dimensional nonlinear features, the embedding layer provides a unified input representation while reducing the computational complexity of directly processing high-dimensional input.

            In G2P prediction tasks, high-dimensional nonlinear features typically include thousands of single nucleotide polymorphism markers, which may vary significantly in distribution and importance. The embedding layer enables the model to efficiently process these features by learning appropriate weights to map the input into a fixed representation space while preserving critical information in the genotypic data. This provides a solid foundation for subsequent attention mechanisms and deeper network layers, allowing the model to better capture the high-dimensional nonlinear features.

        \subsubsection{Residual Connections for Preserving Information Flow}
            To mitigate the vanishing gradient problem and improve the stability of deep neural networks, residual connections are incorporated into the model. These connections directly pass the input of a layer to its output by adding it to the layer's transformation result. The formula for residual connections is expressed as follows:
            \begin{align}
                \mathbf{y} = \mathbf{v} + \mathcal{F}(\mathbf{v}, \Theta)
            \end{align}
            where \(\mathbf{y}\) represents the output of the layer, \(\mathbf{v}\) is the input, and \(\mathcal{F}(\mathbf{v}, \Theta)\) is the transformation applied by the layer with trainable parameters \(\Theta\).

            Residual connections allow the model to bypass certain transformations by directly adding the input to the output of the transformation layer. This technique mitigates the degradation problem in deep networks by ensuring that critical information from earlier layers is preserved throughout the network. In this design, the transformation \(\mathcal{F}(\mathbf{x}, \Theta)\) learns residual mappings, focusing on refining or complementing the input data rather than fully transforming it. This simplifies the optimization process for deep networks.

            In G2P prediction, where the input data contains thousands of features, residual connections play a vital role in maintaining the flow of information through the model's layers. By preserving the original input, residual connections help prevent the loss of crucial genotypic information while enabling the model to learn complex transformations necessary for phenotype prediction. This structure ensures that deep networks can  capture the high-dimensional nonlinear features without overfitting or convergence issues, making them particularly suitable for high-dimensional and complex G2P prediction tasks.

    \subsection{Constraints}
        The study focuses on the G2P prediction task using genomic data obtained from the IRRI Rice299 dataset \citep{li2024trg2p}. The analysis is conducted on 299 samples, with 5,000 SNP markers selected after preprocessing to ensure computational feasibility. The data is split into training, validation, and test sets using a 70:15:15 ratio, ensuring consistent proportions across all experiments. Random sampling is applied during the data split to maintain the representativeness of the dataset.

        Additionally, the dataset was transformed from nucleotide genotype coding ('A', 'C', 'T', 'G') to numeric coding (1, 0, -1 for class I homozygotes, heterozygotes, and class II homozygotes) to facilitate statistical analysis. Meanwhile, phenotypic data was represented as continuous numerical values for each trait, enabling a seamless integration with the genotypic dataset for predictive modeling. 

        The model assumes that complex nonlinear relationships exist between genotypes and phenotypes, requiring advanced architectures to capture these interactions. MLFformer employs attention mechanisms to extract long-range dependencies from high-dimensional genomic data and integrates MLP structures to model intricate nonlinear patterns effectively. Additionally, MLFformer uses Fast Attention to approximate self-attention mechanisms efficiently on high-dimensional data.

        Hyperparameter optimization is performed using the Tree-structured Parzen Estimator method. The model’s predictive accuracy is comprehensively evaluated using multiple metrics, including Mean Absolute Percentage Error (MAPE), Mean Absolute Error (MAE), Mean Absolute Deviation Percentage (MADP), Mean Squared Error (MSE), and Root Mean Squared Error (RMSE). These metrics provide a robust and systematic assessment of performance.
    
    \subsection{Procedures and Training}
        The research adopts a simulation-based modeling approach to investigate G2P prediction. The study design includes four key modules: data preprocessing, model training, model validation, and performance evaluation, ensuring a structured and reproducible workflow. 
        
        To begin with, the IRRI Rice299 dataset, consisting of 299 samples and 5,000 SNP markers, is used for G2P prediction. SNP markers serve as predictors, and phenotypes represent target variables. The dataset is divided into training, validation, and test sets in a 70:15:15 ratio through random sampling. This division maintains consistent representation across experiments and supports robust model evaluation.
       
        Subsequently, the model uses the training set to minimize the MSE loss function, enabling the learning of relationships between SNP markers and phenotypes. The Adam optimizer updates model parameters efficiently, facilitating convergence during training.
        
        Next, the validation set monitors model performance and ensures generalization by applying an early stopping strategy. Validation loss is calculated at the end of each epoch to evaluate performance. Training terminates early if validation loss does not improve within a predefined patience period, avoiding overfitting.
        
\begin{figure}[h]\centering
	\includegraphics[width=\linewidth]{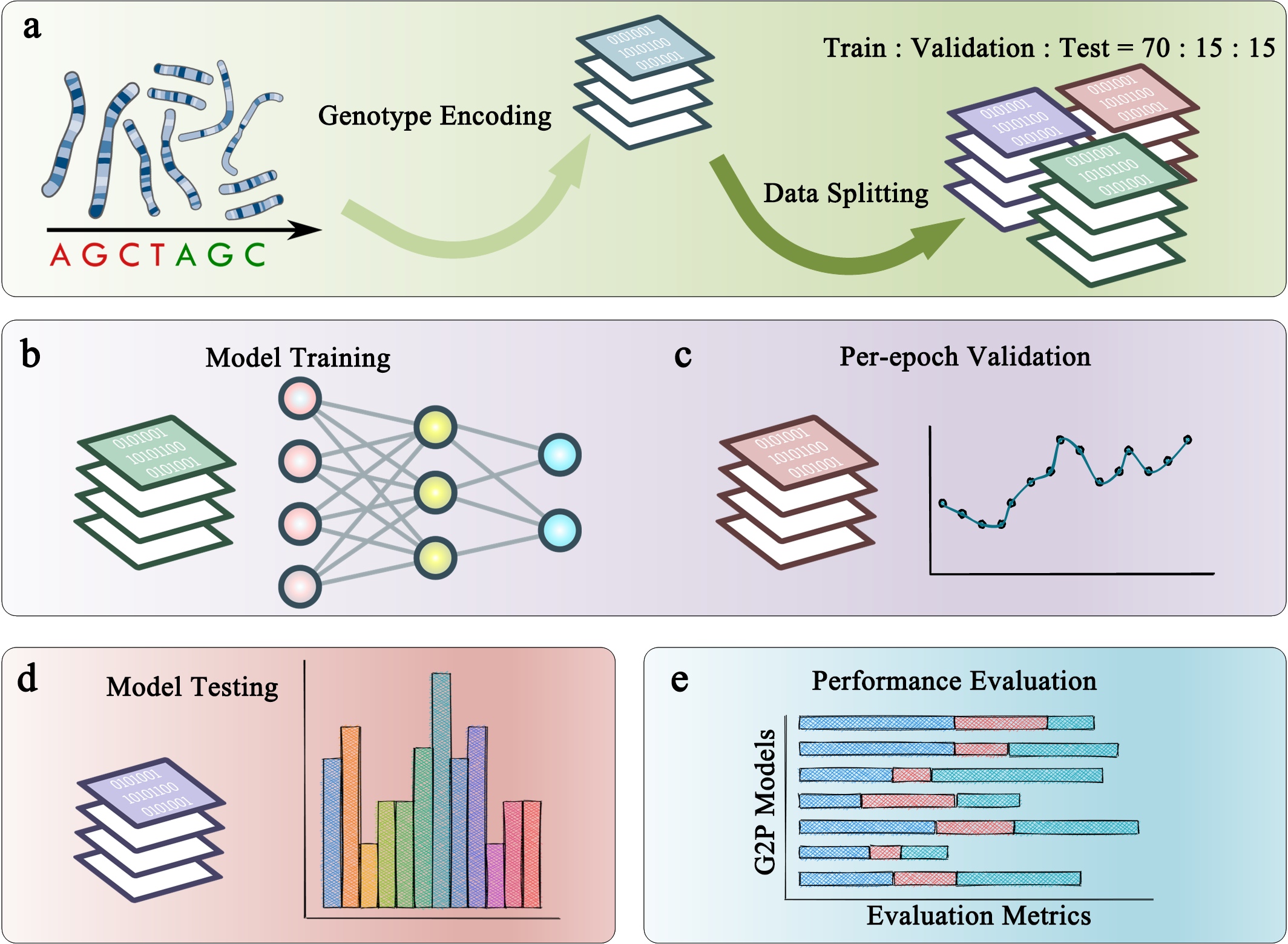}
	\caption{The experimental workflow.} 
    \label{Procedures}
    \end{figure}

        As the final step, the trained model is evaluated on the test set to assess generalization across phenotypes. Evaluation metrics, including MAPE, MAE, MADP, MSE, and RMSE, are applied to analyze predictive accuracy comprehensively. These metrics assess both overall performance and sensitivity to extreme values, providing a detailed understanding of the model's capabilities. The experimental workflow is illustrated in the Fig.~\ref{Procedures}.

\section{Experimental Results and Discussion}
\label{Experimental Results and Discussion}

    \subsection{Experimental Data}
    The Rice299 dataset was obtained from the irrigated rice breeding program of the International Rice Research Institute \citep{li2024trg2p}. 
            
    To analyze the genetic diversity across all rice lines, genotyping by sequencing was employed to identify and genotype single nucleotide polymorphisms (SNPs). To ensure the quality of the dataset, SNPs with call rates below 90\% and monomorphic markers were excluded after imputation, resulting in a filtered dataset containing 73,147 SNPs. To further simplify the dataset and facilitate its usage, a binning strategy was then implemented to reduce the number of SNP markers from 73,147 to 5,000, ensuring a manageable marker set for downstream analysis. 
            
    Phenotypic data, including grain length (GL), grain width (GW), flag leaf area (LA), plant height (PH), panicle number (PnN), yield (YLD), and yield per plant (YPP), were extracted from historical breeding trial records conducted at a single location in Los Baños between 2009 and 2012. By merging the genotypic and phenotypic data, 299 accessions were retained for the study. The resulting dataset provided a reliable foundation for evaluating predictive modeling approaches in rice breeding research.

    \subsection{Compared Models}
        Our experiments selected eight Transformer-based models as compared models: Informer, Longformer, Vanilla Transformer, Transformer-XL, Nonstationary Transformer, Sparseformer, Temporal Fusion Transformer, and iTransformer. These models were chosen for their demonstrated ability to handle sequence-based data, innovative architectural designs adapted for diverse tasks, and effectiveness in addressing challenges in processing high-dimensional genomic datasets. The selection of Transformer-based architectures was motivated by their proven capability to capture long-range dependencies, model nonlinear relationships, and adapt to both single-task and multitask learning scenarios, making them well-suited for genotype-to-phenotype prediction tasks.
        \subsubsection{Informer Based on Sparse Mechanism \citep{gong2022load}}
            Informer was designed for long-sequence time series forecasting, emphasizing efficiency and accuracy. The ProbSparse self-attention mechanism reduced the complexity of traditional self-attention from \(O(L^2)\) to \(O(L \log L)\). The mechanism selected dominant queries with higher sparsity for attention computation, improving computational efficiency while preserving key dependencies. Informer also applied a self-attention distillation operation to downsample intermediate features, reducing sequence length and accelerating inference. The design achieved significant recognition for efficiently handling large-scale data and long sequences.

        \subsubsection{Longformer Based on Sliding Window Mechanism \citep{beltagy2020longformer}}
            Longformer addressed the limitations of traditional Transformers in handling long sequences by introducing a sliding window self-attention mechanism. The sliding window reduced computational complexity from \(O(L^2)\) to \(O(L)\), making the model scalable for long-sequence tasks. Global attention was incorporated to allow connections between distant tokens, ensuring critical information was captured across sequences. The mechanism demonstrated strong performance in analyzing extensive genomic datasets.

        \subsubsection{Vanilla Transformer Based on Standard Structure \citep{vaswani2017attention}}
            Vanilla Transformer represented the foundational architecture of Transformer models and was originally designed for sequence-to-sequence tasks. The architecture employed a standard multi-head self-attention mechanism and feed-forward neural networks to model dependencies within sequences. Computational complexity at \(O(L^2)\) limited scalability for very long sequences. Vanilla Transformer served as the baseline model for comparisons in G2P prediction tasks.

        \subsubsection{Transformer-XL Based on Segment-Level Recurrence \citep{min2024application}}
            Transformer-XL extended the Transformer architecture by introducing a segment-level recurrence mechanism, enabling the reuse of hidden states across segments. This approach improved the ability to capture long-term dependencies while reducing memory usage. Relative position encoding provided better representation of positional information across segments. These innovations allowed Transformer-XL to handle long sequences efficiently in genomic data analysis.

        \subsubsection{Nonstationary Transformer Based on Nonstationary Characteristics \citep{wan2024tcdformer}}
            Nonstationary Transformer was specifically designed to address challenges posed by nonstationary time-series data. Specialized modules modeled and adapted to varying statistical properties in the data, such as shifting distributions or changing correlations. The architecture effectively captured dynamic patterns, making it suitable for tasks with nonstationary characteristics. Adaptability in scenarios involving variable data properties provided a distinct advantage.

        \subsubsection{Sparseformer Based on Sparse Attention \citep{huang2024sparse}}
            Sparseformer improved efficiency by incorporating sparsity into the attention mechanism. Focusing on a subset of relevant tokens reduced computational complexity while maintaining predictive accuracy. The sparse attention mechanism was particularly effective for high-dimensional data, where sparsity helped manage computational burden. Sparseformer balanced computational efficiency and accuracy, making it suitable for large-scale genomic datasets.

        \subsubsection{Temporal Fusion Transformer Based on Temporal Integration \citep{wu2022interpretable}}
            Temporal Fusion Transformer (TFT) was tailored for multi-horizon time-series forecasting. Static covariates, known inputs, and observed inputs were combined to build a unified representation. Interpretable attention mechanisms, including static and temporal attention, highlighted the key drivers of predictions. Gating mechanisms selectively utilized relevant features to enhance efficiency and explainability. Temporal Fusion Transformer provided a framework for applications requiring accurate predictions and interpretability.

        \subsubsection{iTransformer Based on Enhanced Efficiency \citep{pei2025tackling}}
            iTransformer was built on the core Transformer architecture and introduced enhancements for scalability and efficiency. The improved attention mechanism reduced computational complexity while retaining critical dependencies. Specialized modules improved performance for high-dimensional genomic data and multitask learning scenarios. The design proved effective for G2P prediction tasks, particularly in addressing efficiency and adaptability requirements.
            
\subsection{Experimental Hyperparameters}
    The main hyperparameters in the experiment are $d$, $h$, $l$, $p$, and $\eta$. As shown in the Table~\ref{Hyperparameter Settings for compared}, $d$ specifies the dimensionality of the model's hidden states. $h$ determines the number of attention heads in the multi-head attention mechanism. $l$ sets the depth of the model, influencing its capacity. $p$ helps prevent overfitting by randomly setting some input units to zero during training. The $\eta$ controls the step size during optimization, affecting the speed of convergence.

    \begin{table}[h]
    \centering
    \caption{Hyperparameter Settings for compared}
    \label{Hyperparameter Settings for compared}
    \begin{tabular}{lccccccc}
    \toprule
         Model & $d$ & $h$ & $l$ & $p$ & $\eta$ \\
    \midrule
        MLFformer
                    &  256 
                    &  8
                    &  8
                    &  0.1
                    &  0.0001 \\
                Informer
                    &  512 
                    &  8
                    &  3
                    &  0.1
                    &  0.0001 \\
                Longformer
                    &  256 
                    &  4
                    &  8
                    &  0.1
                    &  0.0001 \\
               Vanilla Transformer
                    &  256 
                    &  8
                    &  8
                    &  0.1
                    &  0.0001 \\
                Transformer-XL
                    &  256 
                    &  8
                    &  8 
                    &  0.1
                    &  0.0001 \\
                Nonstationary Transformer
                    &  32 
                    &  2
                    &  2 
                    &  0.1
                    &  0.0001 \\
                Sparseformer
                    &  256 
                    &  8
                    &  4
                    &  0.1
                    &  0.0001 \\
               Temporal Fusion Transformer
                    &  256 
                    &  4
                    &  4
                    &  0.1
                    &  0.0001\\
              iTransformer
                    &  256 
                    &  8 
                    &  8
                    &  0.1 
                    &  0.0001  \\
    \bottomrule
    \end{tabular}
    \end{table}

\subsection{Evaluation Metrics}  
    \subsubsection{MAPE}
     MAPE measures the average percentage error between predicted (\(\hat{\mathbf{y}}_{mw}\)) and observed (\(\mathbf{y}_{mw}\)) phenotype values. In the formula, \(m\) represents the sample index (ranging from 1 to 299), and \(w\) represents the phenotype index (ranging from 1 to 7, corresponding to the 7 different phenotypes). MAPE is defined as:
    \begin{align}
        \mathrm{MAPE} = \frac{1}{n} \sum_{i=1}^{n} \left| \frac{\hat{\mathbf{y}}_{mw} - \mathbf{y}_{mw}}{\mathbf{y}_{mw}} \right| \times 100
    \end{align}
    where \(n\) is the number of samples. MAPE quantifies the average prediction error as a percentage of the true values.
     
     \subsubsection{MAE} 
     MAE measures the average absolute deviation between predicted (\(\hat{\mathbf{y}}_{mw}\)) and observed (\(\mathbf{y}_{mw}\)) phenotype values. MAE is defined as:
    \begin{align}
    \mathrm{MAE} = \frac{1}{n} \sum_{i=1}^{n} \left| \hat{\mathbf{y}}_{mw} - \mathbf{y}_{mw} \right|
    \end{align}
    MAE quantifies the average absolute deviation from the true values.

     \subsubsection{MADP} 
     MADP measures the average absolute deviation between predicted (\(\hat{\mathbf{y}}_{mw}\)) and observed (\(\mathbf{y}_{mw}\)) values, expressed as a percentage of the mean of the observed values (\(\overline{\mathbf{y}}_w\)). MADP is defined as:
    \begin{align}
    \mathrm{MADP} = \frac{1}{n} \sum_{i=1}^{n} \left| \frac{\hat{\mathbf{y}}_{mw} - \mathbf{y}_{mw}}{\overline{\mathbf{y}}_w} \right| \times 100
    \end{align}
    where \(\overline{\mathbf{y}}_w = \frac{1}{n} \sum_{i=1}^{n} \mathbf{y}_{mw}\). MADP quantifies the average absolute deviation as a percentage-based representation.

    \subsubsection{MSE} 
    MSE measures the average squared deviation between predicted (\(\hat{\mathbf{y}}_{mw}\)) and observed (\(\mathbf{y}_{mw}\)) phenotype values. MSE is defined as:
    \begin{align}
    \mathrm{MSE} = \frac{1}{n} \sum_{i=1}^{n} (\hat{\mathbf{y}}_{mw} - \mathbf{y}_{mw})^2
    \end{align}
    MSE quantifies the average magnitude of squared residuals, penalizing larger errors more heavily.

    \subsubsection{RMSE} 
    RMSE measures the square root of the average squared deviation between predicted (\(\hat{\mathbf{y}}_{mw}\)) and observed (\(\mathbf{y}_{mw}\)) phenotype values. RMSE is defined as:
    \begin{align}
    \mathrm{RMSE} = \sqrt{\frac{1}{n} \sum_{i=1}^{n} (\hat{\mathbf{y}}_{mw} - \mathbf{y}_{mw})^2}
    \end{align}
    RMSE quantifies the average magnitude of residuals and is more sensitive to outliers due to the squaring of the error terms.

\subsection{ Experimental Results}
    A series of experiments is designed to evaluate the performance and adaptability of MLFformer in genotype-phenotype prediction tasks. First, MLFformer is compared with other Transformer-based models to establish its baseline performance. Next, the effect of incorporating an MLP structure into various models is examined to determine whether similar improvements can be achieved. Then, the experiments further test the predictive performance of MLFformer under multivariate prediction scenarios. Finally, hyperparameter optimization is carried out to improve MLFformer’s prediction accuracy. These experiments together form a framework for evaluating Transformer-based models and demonstrate MLFformer’s potential as a reliable tool for genomic studies. 

    \subsubsection{\textbf{Genotype-Phenotype Prediction Performance Comparison.} }
        MLFformer demonstrates significant advantages in G2P prediction tasks compared to the compared models. Performance is assessed using multiple metrics, including MAPE, MAE, MADP, MSE, and RMSE, across seven phenotypes (GL, GW, LA, PH, PnN, YLD, YPP). The results for each metric and phenotype are presented in detail in the corresponding tables.
        
        Table~\ref{tab:mape1} shows that MLFformer achieves the lowest MAPE values for GL, GW, PnN, YLD, and YPP, highlighting its capacity to model complex genotype-phenotype relationships. Table~\ref{tab:mae1} reports that MLFformer performs best for GL, GW, PnN, and YPP. Table~\ref{tab:madp1} indicates that MLFformer demonstrates consistent performance across all seven phenotypes, with particularly strong results for GL, GW, PnN, YLD, and YPP. Furthermore, Table~\ref{tab:mse1} shows that MLFformer achieves lower MSE values for GL, GW, and PnN, while Table~\ref{tab:rmse1} confirms its robust predictive performance across most phenotypes, especially GL, GW, and PnN. Longformer, Informer, and Nonstationary Transformer demonstrate good performance on certain metrics, while Sparseformer and ITransformer exhibit lower predictive accuracy on this dataset.

        In comparison to MLFformer, Longformer and Informer perform well on specific metrics. But overall, MLFformer demonstrates better performance. While the models handle long sequences, they still exhibit limitations in addressing high-dimensional nonlinear features. In contrast, Sparseformer and iTransformer exhibit lower predictive accuracy, reflecting their inability to handle high-dimensional nonlinear features.

        The superior performance of MLFformer can be attributed to its design, which incorporates the Transformer framework, Fast Attention mechanism, and multilayer perceptron structure. The ability of MLFformer to capture complex interactions within data supports its effectiveness in genotype-to-phenotype prediction. This design showcases its potential for application in genomic research and precision breeding.
        
        All in all, the results indicate that MLFformer consistently achieves lower error rates and higher predictive accuracy compared to the other models, emphasizing its effectiveness in G2P prediction tasks. After confirming MLFformer’s strong performance compared to other models, the next step examines a key component of its architecture, the MLP structure. The next investigation focuses on whether the MLP structure can also improve predictive accuracy in the compared models.

    \begin{table}[h]
    \centering
    \caption{Model Performance Comparison: MAPE for Different Phenotypes}
    \label{tab:mape1}
    \begin{tabularx}{\textwidth}{lXXXXXXXXXX}
        \toprule
        & \multicolumn{7}{c}{MAPE (\%) $\downarrow$} \\
        \cmidrule(lr){2-8}
        Model & GL & GW & LA & PH & PnN & YLD & YPP \\
        \midrule
        \textcolor{gold}{MLFformer}
            & \textcolor{gold}{2.46 }  
            & \textcolor{gold}{4.85 }   
            & 7.85   
            & \textcolor{silver}{4.61 }   
            & \textcolor{gold}{3.93  }   
            & \textcolor{gold}{9.17 }   
            & \textcolor{gold}{4.53}  \\
        Informer
            & 2.83   
            & \textcolor{bronze}{5.49}  
            & 7.90   
            & 4.82  
            & 4.23 
            & 9.52 
            & \textcolor{bronze}{4.89}\\
        Longformer
            & \textcolor{silver}{2.73}   
            & 5.58   
            & \textcolor{gold}{7.72}  
            & 4.80 
            & \textcolor{silver}{4.19 }   
            & \textcolor{bronze}{9.22  } 
            & 4.95  \\
        Vanilla Transformer
            & 2.78 
            & 5.59 
            & \textcolor{bronze}{7.74 }   
            & 4.80 
            & 4.22 
            & 9.80   
            & 4.94  \\
        Transformer-XL
            & \textcolor{bronze}{ 2.75  } 
            & 5.60
            & \textcolor{silver}{ 7.73 }  
            & 4.82   
            & \textcolor{bronze}{ 4.19 }  
            & \textcolor{silver}{9.22 }   
            & 4.94 \\
     Nonstationary Transformer
            & 3.59  
            & \textcolor{silver}{5.33 }  
            & 9.13  
            & \textcolor{bronze}{4.73 }  
            & 4.62 
            & 9.63 
            & \textcolor{silver}{4.76 }  \\
       Sparseformer
            & 7.49   
            & 6.59 
            & 10.56 
            & 7.72 
            & 9.27   
            & 11.43 
            & 5.63  \\
        Temporal Fusion Transformer
            & 3.43  
            & 6.40  
            & 8.52  
            & \textcolor{gold}{4.46}  
            & 4.75  
            & 11.34   
            & 4.93  \\
        iTransformer
            & 6.11
            & 7.67 
            & 9.61 
            & 7.23  
            & 6.53 
            & 10.98  
            & 6.85 \\
    
    \bottomrule
    \end{tabularx}
    
    \footnotetext[1]{\textcolor{gold}{Gold} represents the best result; \textcolor{silver}{silver} represents the second best result; \textcolor{bronze}{bronze} represents the third best result.}
    \footnotetext[2]{The downward arrow ( $\downarrow$ ) signified that a smaller value for the corresponding metric corresponded to superior predictive performance of the model.}
    \end{table}

    \begin{table}[h]
    \centering
    \caption{Model Performance Comparison: MAE for Different Phenotypes}
    \label{tab:mae1}
    \begin{tabularx}{\textwidth}{lXXXXXXXXXX}
        \toprule
        & \multicolumn{7}{c}{MAE $\downarrow$} \\
        \cmidrule(lr){2-8}
        Model & GL & GW & LA & PH & PnN & YLD & YPP \\
        \midrule
        \textcolor{gold}{MLFformer}
        & \textcolor{gold}{0.24} 
        & \textcolor{gold}{1.19} 
        & 1.94 
        & \textcolor{silver}{5.38}
        & \textcolor{gold}{0.51} 
        & \textcolor{silver}{434.44}
        & \textcolor{gold}{1.331} \\
    
    Informer
        & \textcolor{bronze}{0.27}
        & \textcolor{bronze}{1.35} 
        & 1.92 
        & 5.82 
        & \textcolor{silver}{0.53} 
        & 455.84  
        & \textcolor{bronze}{1.44} \\
    Longformer
        & \textcolor{silver}{0.26} 
        & 1.36 
        & \textcolor{gold}{1.90 } 
        & 5.75 
        & \textcolor{bronze}{0.54 } 
        & \textcolor{bronze}{437.08} 
        & 1.46 \\
    Vanilla Transformer
        & 0.27
        & 1.37
        & \textcolor{bronze}{1.92} 
        & 5.75 
        & 0.54 
        & 469.75
        & 1.46  \\
    Transformer-XL
        & 0.27
        & 1.37
        & \textcolor{silver}{1.91} 
        & 5.80
        & 0.54
        & 437.07
        & 1.46\\
    Nonstationary Transformer
        & 0.34
        & \textcolor{silver}{1.34} 
        & 2.24
        & \textcolor{bronze}{5.63} 
        & 0.59
        & \textcolor{gold}{432.17} 
        & \textcolor{silver}{1.40} \\
    Sparseformer
        & 0.63
        & 1.60
        & 2.52
        & 8.81
        & 1.16
        & 559.42
        & 1.65\\
    Temporal Fusion Transformer
        & 0.34
        & 1.60
        & 2.10 
        & \textcolor{gold}{5.29 } 
        & 0.60 
        & 2665.91  
        & 1.47  \\
    iTransformer
        & 0.59 
        & 1.86 
        & 2.34 
        & 8.60  
        & 0.82 
        & 518.55 
        & 1.97  \\
    
    \bottomrule
    \end{tabularx}
    
    \footnotetext[1]{\textcolor{gold}{Gold} represents the best result; \textcolor{silver}{silver} represents the second best result; \textcolor{bronze}{bronze} represents the third best result.}
    \footnotetext[2]{The downward arrow ( $\downarrow$ ) signified that a smaller value for the corresponding metric corresponded to superior predictive performance of the model.}
    \end{table}

    \begin{table}[h]
    \centering
    \caption{Model Performance Comparison: MADP for Different Phenotypes}
    \label{tab:madp1}
    \begin{tabularx}{\textwidth}{lXXXXXXXXXX}
        \toprule
        & \multicolumn{7}{c}{MADP (\%) $\downarrow$} \\
        \cmidrule(lr){2-8}
        Model & GL & GW & LA & PH & PnN & YLD & YPP \\
        \midrule
        \textcolor{gold}{MLFformer}
                & \textcolor{gold}{1.22}  
                & \textcolor{gold}{2.39}   
                & 3.93  
                & \textcolor{silver}{2.24}   
                & \textcolor{gold}{1.98}   
                & \textcolor{gold}{4.54}   
                & \textcolor{gold}{2.27}  \\
            Informer
                & 1.42  
                & \textcolor{bronze}{2.72}   
                & \textcolor{bronze}{3.88}  
                & 2.42  
                & 2.17  
                & 4.76  
                & \textcolor{bronze}{2.46}  \\
            Longformer
                & \textcolor{silver}{1.36}   
                & 2.75  
                & \textcolor{gold}{3.86}  
                & 2.39  
                & \textcolor{silver}{2.12}   
                & \textcolor{silver}{4.57}   
                & 2.50 \\
           Vanilla Transformer
                & 1.37  
                & 2.75  
                & \textcolor{silver}{3.86}   
                & 2.39  
                & \textcolor{bronze}{2.12}   
                & 4.93  
                & 2.49 \\
            Transformer-XL
                & \textcolor{bronze}{1.37}   
                & 2.75  
                & 3.90  
                & 2.41  
                & 2.13  
                & \textcolor{bronze}{4.58}   
                & 2.50 \\
           Nonstationary Transformer
                & 1.73  
                & \textcolor{silver}{2.67}  
                & 4.54  
                & \textcolor{bronze}{2.36}   
                & 2.31  
                & 4.58  
                & \textcolor{silver}{2.39}  \\
            Sparseformer
                & 3.11  
                & 3.23  
                & 5.07  
                & 3.75  
                & 4.42  
                & 5.95  
                & 2.82 \\
            Temporal Fusion Transformer
                & 1.72  
                & 3.19  
                & 4.26  
                & \textcolor{gold}{2.22}  
                & 2.36  
                & 37.78  
                & 2.49 \\
           iTransformer
                & 2.96  
                & 3.72  
                & 4.73  
                & 3.61  
                & 3.22  
                & 5.46  
                & 3.37 \\

    \bottomrule
    \end{tabularx}
    
    \footnotetext[1]{\textcolor{gold}{Gold} represents the best result; \textcolor{silver}{silver} represents the second best result; \textcolor{bronze}{bronze} represents the third best result.}
    \footnotetext[2]{The downward arrow ( $\downarrow$ ) signified that a smaller value for the corresponding metric corresponded to superior predictive performance of the model.}
    \end{table}

    \begin{table}[h]
    \centering
    \caption{Model Performance Comparison: MSE for Different Phenotypes}
    \label{tab:mse1}
    \begin{tabularx}{\textwidth}{lXXXXXXXXXX}
        \toprule
        & \multicolumn{7}{c}{MSE $\downarrow$}\\
        \cmidrule(lr){2-8}
        Model & GL & GW & LA & PH & PnN & YLD & YPP \\
        \midrule
        \textcolor{gold}{MLFformer}
                & \textcolor{gold}{0.10} 
                & \textcolor{gold}{2.26} 
                & 6.98 
                & \textcolor{bronze}{56.63} 
                & \textcolor{gold}{0.42} 
                & \textcolor{silver}{362638} 
                & \textcolor{silver}{3.41} \\
            Informer 
                & 0.14 
                & \textcolor{silver}{2.69} 
                & \textcolor{bronze}{6.95} 
                & 67.65 
                & 0.47 
                & 388605
                & \textcolor{bronze}{3.54} \\
            Longformer
                & \textcolor{silver}{0.12} 
                & 2.73 
                & \textcolor{gold}{6.92} 
                & 65.35 
                & \textcolor{silver}{0.46} 
                & 366294
                & 3.62 \\
            Vanilla Transformer 
                & 0.14 
                & \textcolor{bronze}{2.73} 
                & 6.97 
                & 65.35 
                & \textcolor{bronze}{0.46} 
                & 405138
                & 3.61 \\
            Transformer-XL
                & \textcolor{bronze}{0.13} 
                & 2.74 
                & \textcolor{silver}{6.93} 
                & 66.71 
                & 0.49 
                & \textcolor{bronze}{366279} 
                & 3.61 \\
            Nonstationary Transformer 
                & 0.21 
                & 2.84 
                & 7.67 
                & \textcolor{silver}{53.13} 
                & 0.53 
                & \textcolor{gold}{313210} 
                & \textcolor{gold}{3.30} \\
            Sparseformer
                & 0.66 
                & 4.23 
                & 10.49 
                & 129.80 
                & 2.03 
                & 505123
                & 4.49 \\
            Temporal Fusion Transformer
                & 0.21 
                & 4.04 
                & 7.77 
                & \textcolor{gold}{47.74} 
                & 0.53 
                & 7388916
                & 4.12 \\
            iTransformer
                & 0.52 
                & 5.42 
                & 9.31 
                & 114.17 
                & 0.98 
                & 447783 
                & 6.82 \\

    \bottomrule
    \end{tabularx}
    
    \footnotetext[1]{\textcolor{gold}{Gold} represents the best result; \textcolor{silver}{silver} represents the second best result; \textcolor{bronze}{bronze} represents the third best result.}
    \footnotetext[2]{The downward arrow ( $\downarrow$ ) signified that a smaller value for the corresponding metric corresponded to superior predictive performance of the model.}
    \end{table}

    \begin{table}[h]
    \centering
    \caption{Model Performance Comparison: RMSE for Different Phenotypes}
    \label{tab:rmse1}
    \begin{tabularx}{\textwidth}{lXXXXXXXXXX}
        \toprule
        & \multicolumn{7}{c}{RMSE $\downarrow$} \\
        \cmidrule(lr){2-8}
        Model & GL & GW & LA & PH & PnN & YLD & YPP \\
        \midrule
       \textcolor{gold}{MLFformer}
                & \textcolor{gold}{0.31} 
                & \textcolor{gold}{1.50} 
                & 2.64 
                & \textcolor{bronze}{7.53} 
                & \textcolor{gold}{0.65} 
                & \textcolor{silver}{602.19} 
                & \textcolor{silver}{1.85} \\
            Informer
                & 0.37 
                & \textcolor{silver}{1.64} 
                & \textcolor{bronze}{2.64} 
                & 8.22 
                & 0.69 
                & 623.28 
                & \textcolor{bronze}{1.88} \\
            Longformer
                & \textcolor{silver}{0.34} 
                & 1.66 
                & \textcolor{gold}{2.62} 
                & 8.08 
                & \textcolor{silver}{0.68} 
                & 605.22 
                & 1.90 \\
            Vanilla Transformer
                & 0.36 
                & \textcolor{bronze}{1.65} 
                & 2.65 
                & 8.08 
                & \textcolor{bronze}{0.68} 
                & 634.02 
                & 1.90 \\
            Transformer-XL
                & \textcolor{bronze}{0.35} 
                & 1.66 
                & \textcolor{silver}{2.63} 
                & 8.17 
                & 0.70 
                & \textcolor{bronze}{605.21} 
                & 1.90 \\
            Nonstationary Transformer
                & 0.46 
                & 1.68 
                & 2.75 
                & \textcolor{silver}{7.27} 
                & 0.72 
                & \textcolor{gold}{555.64} 
                & \textcolor{gold}{1.81} \\
            Sparseformer
                & 0.75 
                & 2.05 
                & 3.21 
                & 11.29 
                & 1.40 
                & 704.06 
                & 2.12 \\
            Temporal Fusion Transformer 
                & 0.46 
                & 2.01 
                & 2.74 
                & \textcolor{gold}{6.89} 
                & 0.72 
                & 2717.35 
                & 2.02 \\
           iTransformer
                & 0.72 
                & 2.33 
                & 3.05 
                & 10.69 
                & 0.99 
                & 669.17 
                & 2.61 \\

    \bottomrule
    \end{tabularx}
    
    \footnotetext[1]{\textcolor{gold}{Gold} represents the best result; \textcolor{silver}{silver} represents the second best result; \textcolor{bronze}{bronze} represents the third best result.}
    \footnotetext[2]{The downward arrow ( $\downarrow$ ) signified that a smaller value for the corresponding metric corresponded to superior predictive performance of the model.}
    \end{table}

\subsubsection{\textbf{Impact of MLP Structure on the Performance of Other Transformer Models.} }

        The analysis investigates the potential of introducing the MLP structure from MLFformer into the compared models to enhance prediction accuracy. The study further evaluates whether MLFformer maintains superior prediction accuracy compared to models improved by the addition of the MLP structure.

\begin{figure}[H]\centering
	\includegraphics[width=0.8\linewidth]{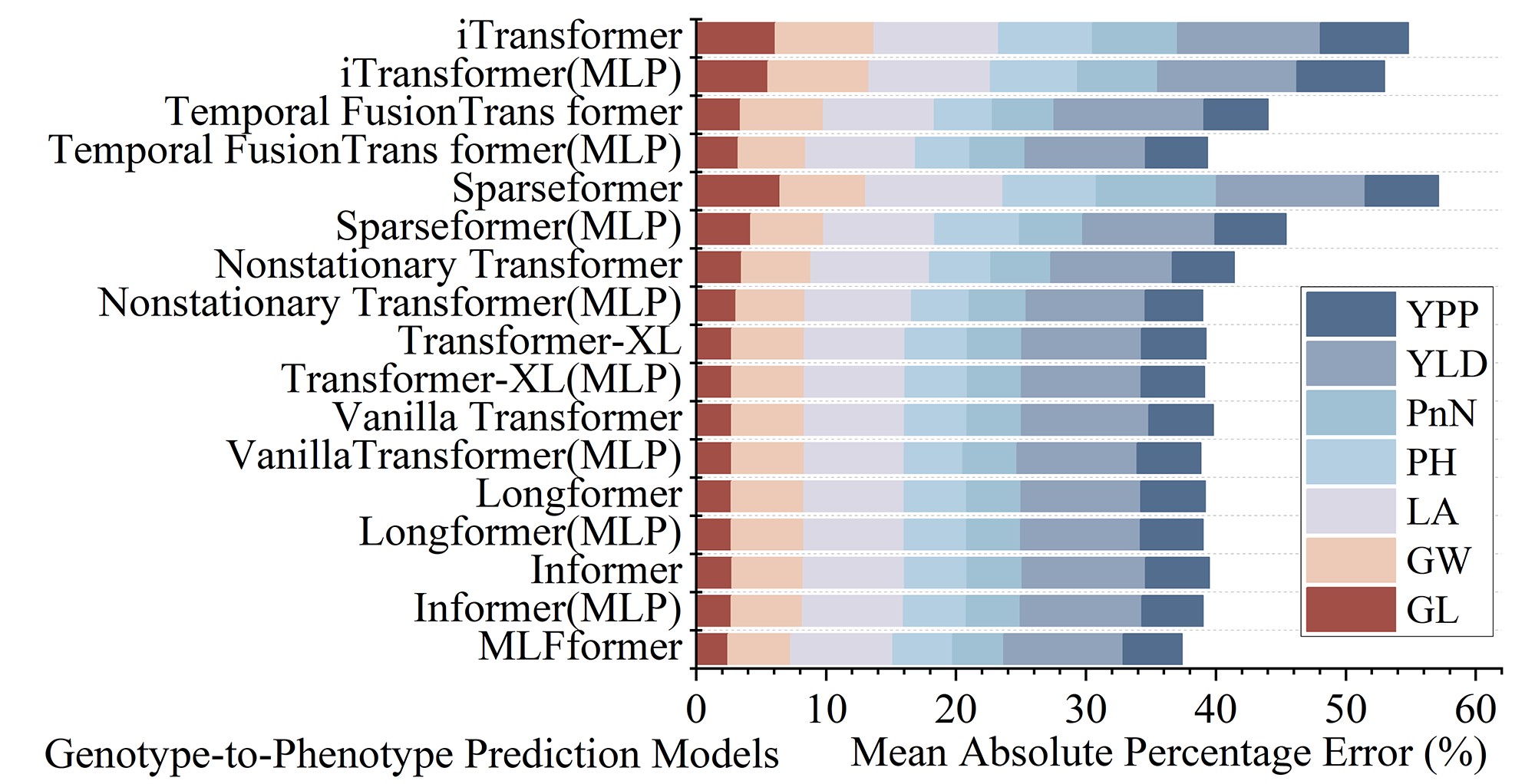}
	\caption{Comparison of Genotype-to-Phenotype Prediction Models Based on MAPE: With and Without MLP} 
    \label{pic:mape2}
    \end{figure}

\begin{figure}[H]\centering
	\includegraphics[width=0.8\linewidth]{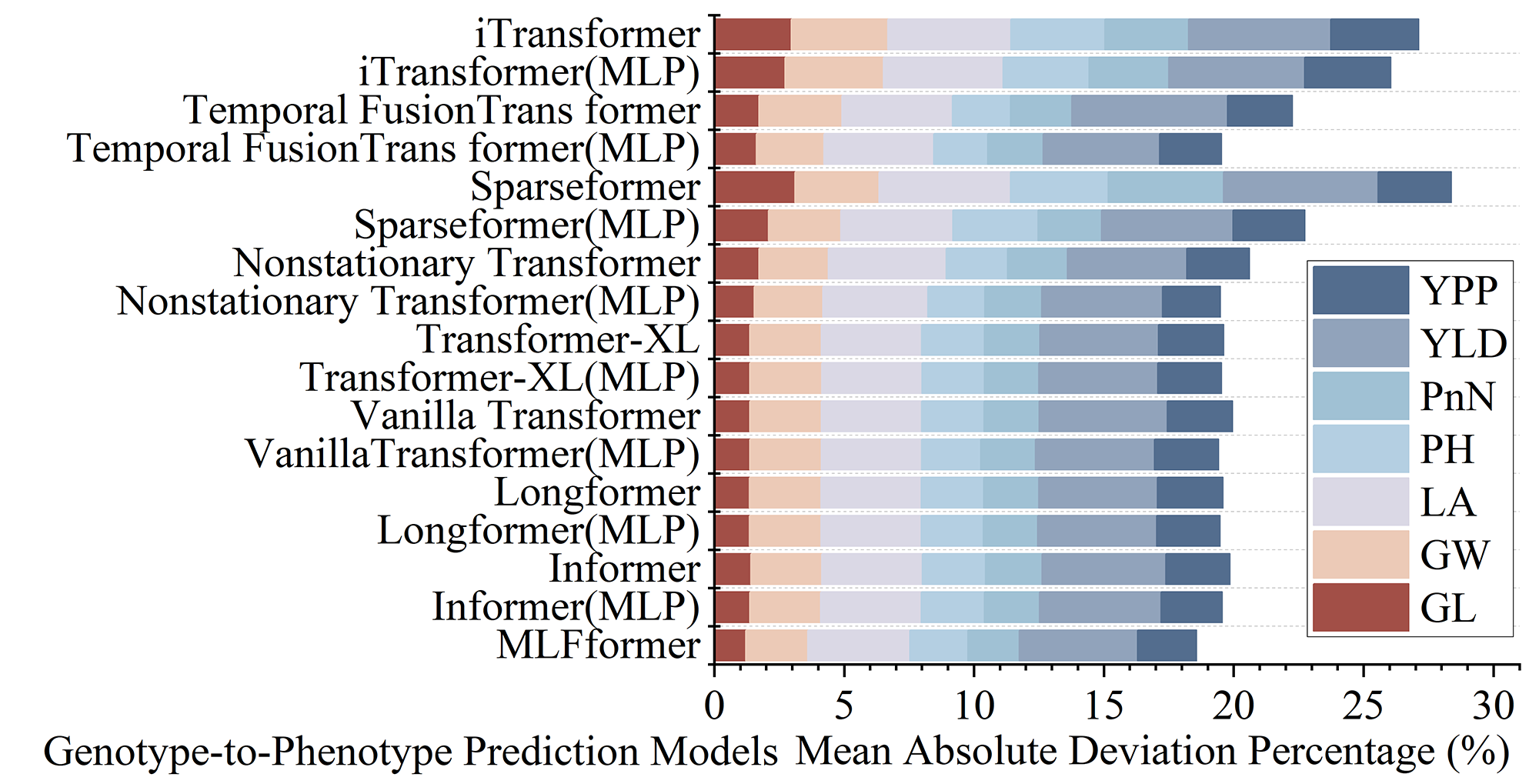}
	\caption{Comparison of Genotype-to-Phenotype Prediction Models Based on MADP: With and Without MLP} 
    \label{pic:madp2}
    \end{figure}

\begin{figure}[H]\centering
	\includegraphics[width=\linewidth]{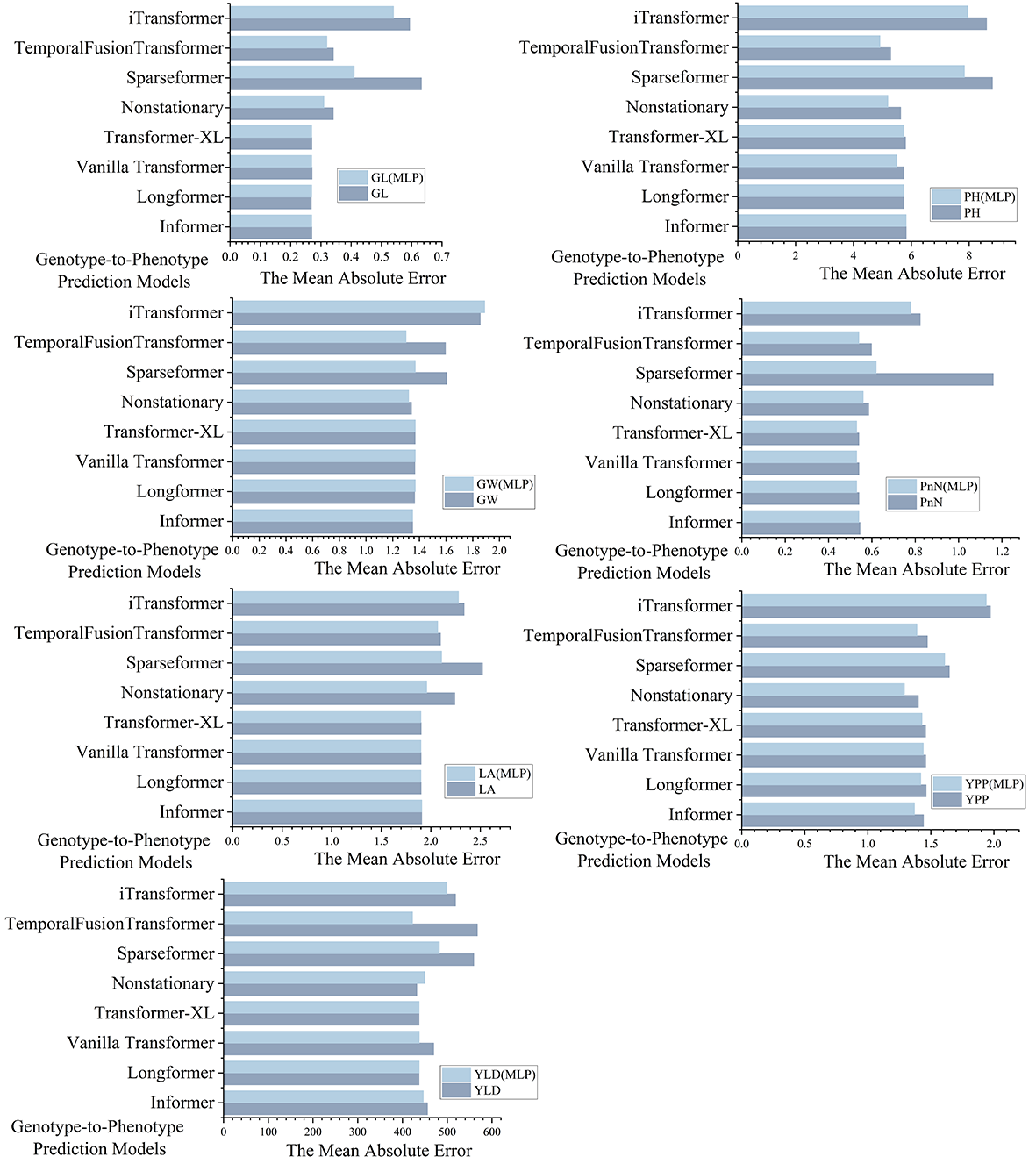}
	\caption{Comparison of Genotype-to-Phenotype Prediction Models Based on MAE: With and Without MLP} 
    \label{pic:mae2}
    \end{figure}

\begin{figure}[H]\centering
	\includegraphics[width=\linewidth]{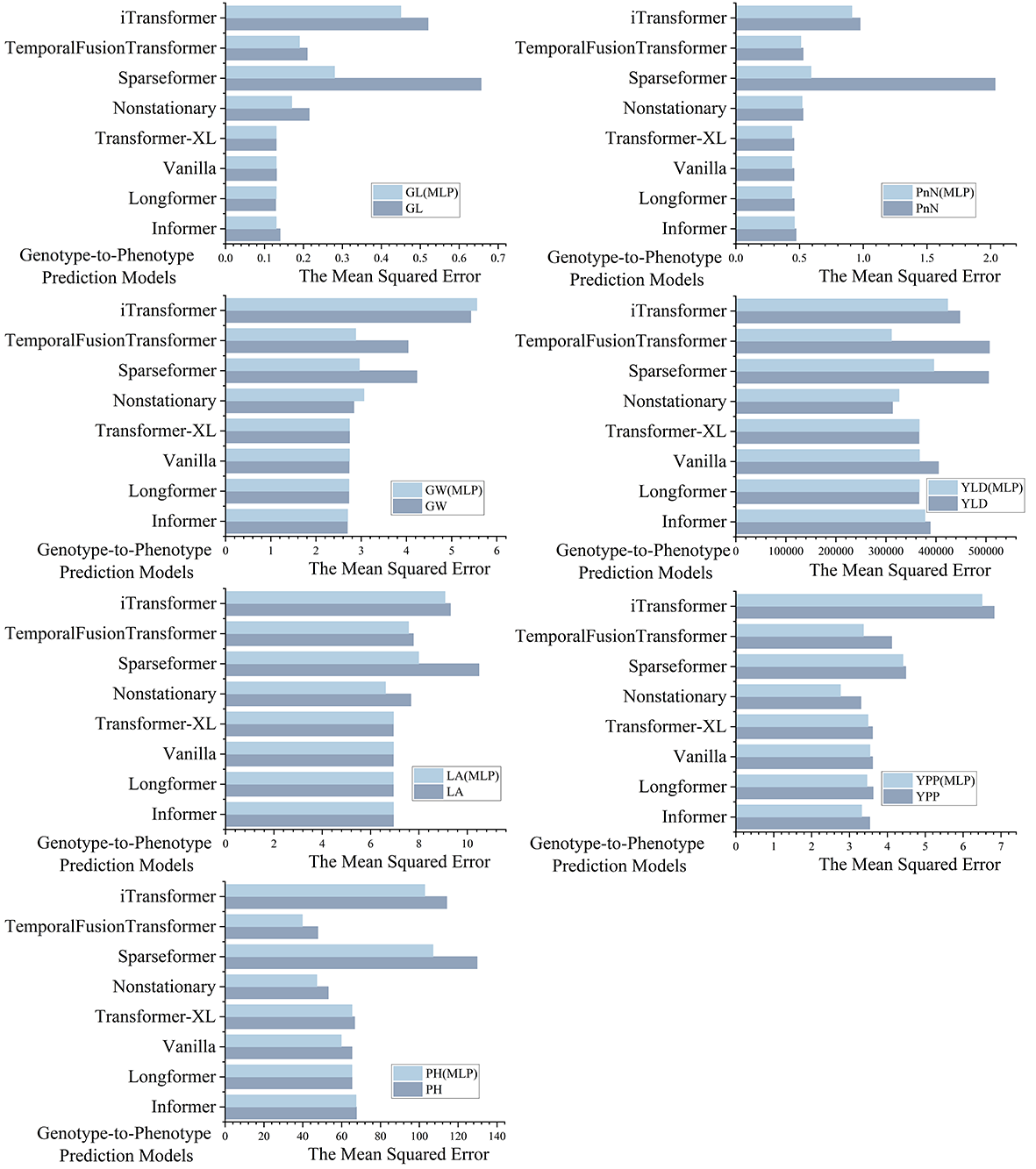}
	\caption{Comparison of Genotype-to-Phenotype Prediction Models Based on MSE: With and Without MLP} 
    \label{pic:mse2}
    \end{figure}

\begin{figure}[H]\centering
	\includegraphics[width=\linewidth]{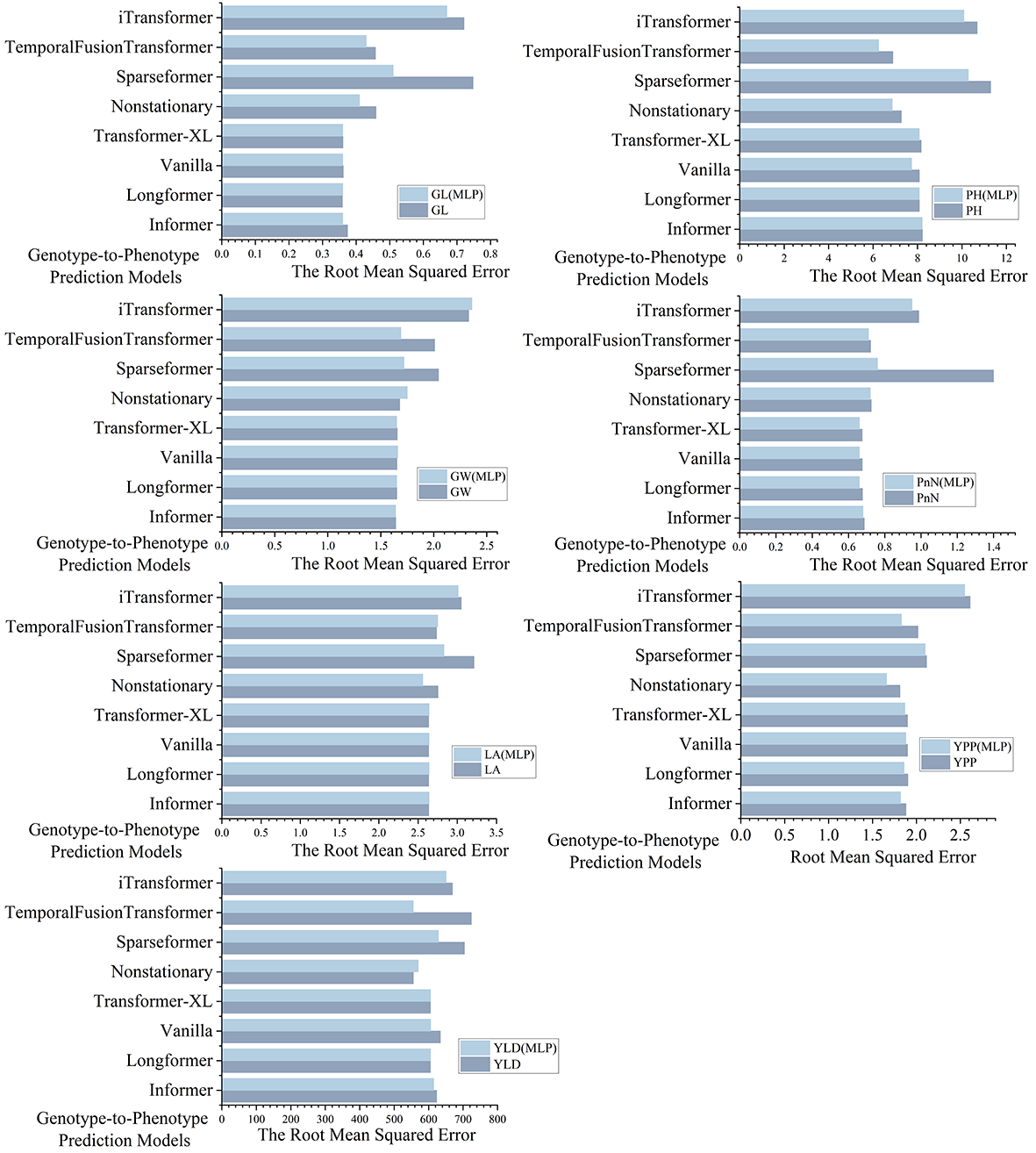}
	\caption{Comparison of Genotype-to-Phenotype Prediction Models Based on RMSE: With and Without MLP} 
    \label{pic:rmse2}
    \end{figure}

        As shown in Fig.~\ref{pic:mape2} and \ref{pic:madp2}, the addition of the MLP structure reduces MAPE and MADP values in most models. Fig.~\ref{pic:mae2}, \ref{pic:mse2}, and \ref{pic:rmse2} indicate that the MLP structure also lowers MSE, MAE, and RMSE values. Temporal Fusion Transformer and Sparseformer achieve the most significant improvements, with Sparseformer showing over a 20\% increase in prediction accuracy for certain phenotypes and metrics. Other models also benefit from the incorporation of the MLP structure.
        
        The results highlight the role of the MLP structure in improving prediction accuracy for G2P tasks. While the MLP structure improves the compared models, MLFformer maintains higher prediction accuracy and achieves better performance than these models. After analyzing the impact of the MLP structure on the compared models, the investigation focuses on evaluating MLFformer’s performance under multitask learning conditions. The step assesses whether MLFformer retains its prediction accuracy when predicting multiple phenotypes. 
        
    \subsubsection{\textbf{Evaluation of MLFformer and Compared Models in Multivariate Prediction Scenarios} }

        From the Table~\ref{tab:duo3}, it can be observed that even under multivariate prediction scenarios, MLFformer maintained the best predictive performance, demonstrating its robustness and adaptability in G2P prediction. On the other hand, the results also highlight the capability of MLFformer in handling high-dimensional nonlinear features.
        
        MSE and RMSE increase the average error due to large prediction errors in individual phenotypes. Therefore, our experiment uses MAPE, MAE, and MADP to better evaluate overall performance. From the Table~\ref{tab:duo3}, it can be observed that MLFformer demonstrates the best overall performance in multivariate prediction tasks. Specifically, MLFformer achieves the lowest MADP of 2.78 and the lowest MAE of 63.52, showcasing its robustness in handling high-dimensional nonlinear features and its adaptability to multivariate prediction. Although Longformer achieves the lowest MAPE value of 5.59, it falls behind in MADP and MAE. Informer and Transformer-XL show competitive performance in some metrics but do not surpass MLFformer overall. These results confirm the effectiveness of MLFformer for G2P prediction in multivariate scenarios.

        The results indicate that MLFformer demonstrates strong capability in addressing high-dimensional nonlinear features. The model achieves accurate G2P prediction across multiple phenotypes in multivariate scenarios.

        \begin{table}[h]
        \centering
        \setlength{\tabcolsep}{11pt} 
        \caption{Model Performance Comparison: MADP for Different Phenotypes}
        \label{tab:duo3}
        \begin{tabularx}{\textwidth}{lccccccccccccc}
            \toprule
            & \multicolumn{3}{c}{Average metric $\downarrow$} \\
            \cmidrule(lr){2-4}
            Model & MAPE & MAE & MADP\\
            \midrule
            MLFformer (Multivariate Prediction)
                & \textcolor{bronze}{5.76} 
                & \textcolor{gold}{63.52} 
                & \textcolor{gold}{2.78} \\
            Informer (Multivariate Prediction)
                & 5.79 
                & 65.26 
                & \textcolor{silver}{2.79} \\
            Longformer (Multivariate Prediction)
                &  \textcolor{gold}{5.59} 
                &  64.91
                &  5.88\\
            Vanilla Transformer (Multivariate Prediction)
                &  6.58 
                &  301.90
                &  3.79\\
            Transformer-XL (Multivariate Prediction)
                &  \textcolor{silver}{5.73} 
                &  \textcolor{bronze}{64.84}
                &  3.55\\
            Nonstationary Transformer (Multivariate Prediction)
                &  6.32 
                &  67.39
                &  5.90\\
            Sparseformer (Multivariate Prediction)
                &  21.29
                &  131.76
                &  6.99\\
           Temporal Fusion Transformer (Multivariate Prediction)
                &  6.90 
                &  301.90
                &  7.47\\
            iTransformer (Multivariate Prediction)
                &  6.12 
                &  \textcolor{silver}{63.75}
                &  \textcolor{bronze}{3.44}\\
            
        \bottomrule
        \end{tabularx}
        
        \footnotetext[1]{\textcolor{gold}{Gold} represents the best result; \textcolor{silver}{silver} represents the second best result; \textcolor{bronze}{bronze} represents the third best result.}
        \footnotetext[2]{The downward arrow ( $\downarrow$ ) signified that a smaller value for the corresponding metric corresponded to superior predictive performance of the model.}
        \end{table}

\subsubsection{Impact of Hyperparameter Optimization on MLFformer's Prediction Performance.} 
        The experiment optimizes the hyperparameters of MLFformer using Bayesian optimization based on the Tree-structured Parzen Estimator method. This optimization focuses on improving the predictive performance of MLFformer in G2P prediction tasks. The results indicate that the optimized hyperparameters increase the prediction accuracy of the model across multiple metrics and phenotypes.

\begin{figure}[h]\centering
	\includegraphics[width=0.9\linewidth]{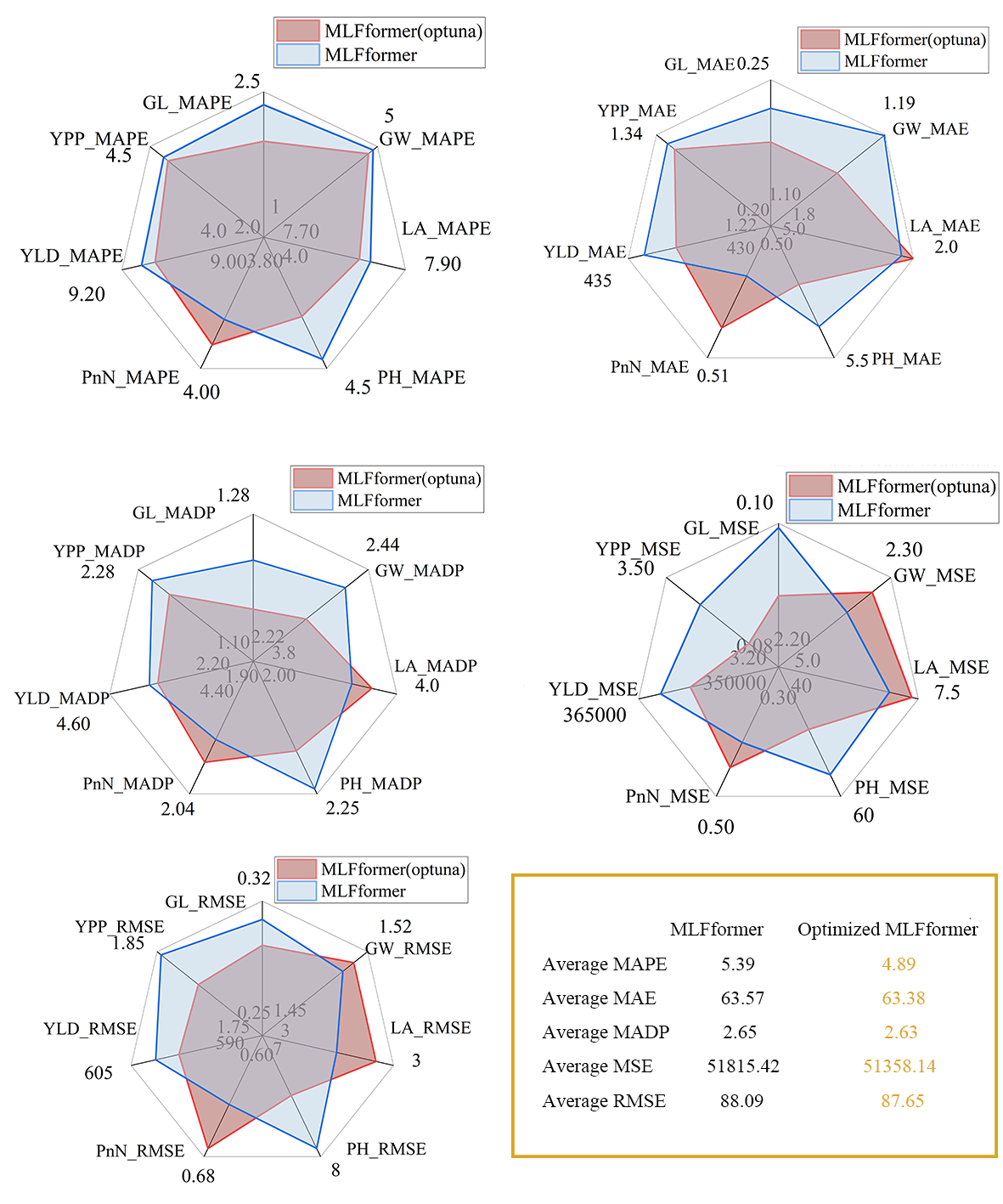}
	\caption{Performance Comparison of MLFformer and Optimized MLFformer Across Multiple Metrics} 
    \label{pic:all4}
    \end{figure}

        The hyperparameters obtained through this process are presented in Table~\ref{tab:hyperparameter5}. To evaluate the impact of the optimization, the performance of the optimized MLFformer is compared with the original version. As shown in Fig.~\ref{pic:all4}, the optimized MLFformer achieves higher accuracy across all major metrics. Specifically, the comprehensive MAPE decreases from 37.78 to 34.24, MAE from 445.02 to 443.70, MADP from 18.60 to 18.41, MSE from 362,708.23 to 359,507.81, and RMSE from 616.66 to 613.58.The performance of the optimized MLFformer is further analyzed across different phenotypes. Improvements are observed for most phenotypes, with GL, PH, YLD, and YPP showing consistent gains across all metrics.

        The experiment results show that the Bayesian optimization process successfully enhances the predictive performance of the optimized MLFformer. The improvements make the model more effective for G2P prediction tasks.

        \begin{table}[h]
            \setlength{\tabcolsep}{8pt} 
            \caption{Hyperparameter Optimization for MLFformer}
            \label{tab:hyperparameter5}
            \begin{tabularx}{\textwidth}{lccccccc}
            \toprule
                    Hyperparameters & GL & GW & LA & PH & PnN & YLD & YPP\\
            \midrule
                    $d$
                        & 320 
                        & 128 
                        & 448 
                        & 192 
                        & 448 
                        & 248 
                        & 262 \\
                    $h$
                        & 4 
                        & 8 
                        & 8 
                        & 4 
                        & 45 
                        & 8 
                        & 8 \\
                    $l$
                        & 4 
                        & 5 
                        & 4 
                        & 12 
                        & 40 
                        & 6 
                        & 8 \\
                    $p$
                        & 0.2856 
                        & 0.4816 
                        & 0.0278 
                        & 0.1306 
                        & 0.1257 
                        & 0.1278 
                        & 0.0469 \\
                    $\eta$
                        & 0.0006 
                        & 0.0056 
                        & 0.0018 
                        & 0.0024 
                        & 0.0001 
                        & 0.0001 
                        & 0.0001 \\
            \bottomrule
            \end{tabularx}
        \end{table}

\subsection{discussion}
    To begin with, the experimental results highlight the effectiveness of MLFformer in addressing the challenges of high-dimensional nonlinear features and improving G2P prediction accuracy. The comparative evaluation with other Transformer-based models demonstrates that MLFformer achieves superior performance across multiple metrics and phenotypes. The performance can be attributed to its ability to capture dependencies within high-dimensional nonlinear features through the Transformer framework. Additionally, the Fast Attention mechanism reduces computational complexity, enabling efficient processing of high-dimensional nonlinear features without sacrificing predictive accuracy.

    Subsequently, the incorporation of the MLP structure contributes to modeling high-dimensional nonlinear features in genomic data. The design allows MLFformer to handle high-dimensional nonlinear features more effectively than models without MLP integration. The experiments examining the addition of MLP structures to other models suggest that this enhancement is critical for improving predictive performance in G2P tasks.

    Then, MLFformer also demonstrates robust performance in multivariate prediction scenarios. The scalability and adaptability of MLFformer to predict multiple phenotypes reflect the model's capacity to generalize across diverse predictive tasks. These results highlight the importance of architectural flexibility in addressing high-dimensional nonlinear features.

    Finally, hyperparameter optimization further refines MLFformer’s ability to balance accuracy and computational efficiency. The optimized parameters enhance the model’s ability to learn from high-dimensional nonlinear features. As a result, the improvement minimizes overfitting, as reflected in the model's consistent performance across all phenotypes.

    The findings suggest that MLFformer is well-suited for addressing the challenges of high-dimensional nonlinear features. The innovative combination of Transformer architecture, Fast Attention mechanism, and MLP structure provides a comprehensive solution for tackling high-dimensional nonlinear features, establishing a new method for G2P prediction tasks.

\section{Conclusion}
        In our paper, the experimental results demonstrate that MLFformer outperforms other Transformer-based models in G2P prediction tasks. Specifically, MLFformer achieves superior accuracy in both univariate and multivariate prediction scenarios, highlighting its robustness in handling high-dimensional nonlinear features. The advantage is attributed to its ability to process high-dimensional nonlinear features, which enhances predictive accuracy. Moreover, the incorporation of the Fast Attention mechanism reduces computational complexity, thereby improving scalability for high-dimensional nonlinear features. In addition, the multilayer perceptron structure captures high-dimensional nonlinear features within genomic data, further boosting model performance. Overall, the findings underline MLFformer’s potential as a reliable and efficient framework for advancing G2P prediction.
        
        Our future research will focus on extending the current work by exploring the integration of additional data modalities, such as environmental and soil data, to further improve prediction accuracy. Investigating the interpretive power of MLFformer and its ability to identify critical genomic features can also provide valuable biological insights. Moreover, developing lightweight versions of the model for real-time deployment in resource-constrained settings can enhance its accessibility and usability. The findings in our study contribute to a foundation for advancing G2P prediction methods and serve as a roadmap for future innovations in agricultural genomics.

\bibliography{sn-bibliography}

\end{document}